\documentclass[%
 reprint,
superscriptaddress,
 amsmath,amssymb,
 aps,
]{revtex4-1}

\usepackage{graphicx}
\usepackage{dcolumn}
\usepackage{bm}
\usepackage{color}

\begin{document}

\preprint{APS/123-QED}

\title{Memory-induced transition from a persistent random walk to circular motion for achiral microswimmers}

\author{N Narinder}
\affiliation{Fachbereich Physik, Universit\"at Konstanz, Konstanz, D-78457, Germany}
\author{Clemens Bechinger}
\affiliation{Fachbereich Physik, Universit\"at Konstanz, Konstanz, D-78457, Germany}
\author{Juan Ruben Gomez-Solano}
\altaffiliation[Present address: ]{Instituto de F\'isica, Universidad Nacional Aut\'onoma de M\'exico, D.F. 04510, M\'exico}
\email{r$_$gomez@fisica.unam.mx}
\affiliation{Fachbereich Physik, Universit\"at Konstanz, Konstanz, D-78457, Germany}

\date{\today}

\begin{abstract}

We experimentally study the motion of light-activated colloidal microswimmers in a viscoelastic fluid. We find that, in such a non-Newtonian  environment,
the active colloids undergo an unexpected transition from enhanced angular diffusion to persistent rotational motion above a critical propulsion speed, despite their spherical shape and stiffness. We observe that, in contrast to chiral asymmetric microswimmers, the resulting circular orbits can spontaneously reverse their sense of rotation and exhibit an angular velocity and a radius of curvature that non-linearly depend on the propulsion speed. By means of a minimal non-Markovian Langevin model for active Brownian motion, we show that these non-equilibrium effects emerge from the delayed response of the fluid with respect to the self-propulsion of the particle without counterpart in Newtonian fluids.

\end{abstract} 


\maketitle

Microswimmers are non-equilibrium systems of great current interest due to their ability to convert energy from their liquid surroundings into active motion~\cite{elgeti2015}. 
In contrast to passive and externally driven particles, rotational motion plays a primordial role for such active systems, since their propulsion is strongly determined by a well-defined orientation~\cite{menzel2015}. 
For many motile microorganisms, e.g. flagellated bacteria, their orientation is mainly affected by internal biochemical processes that lead to abrupt changes of otherwise rather straight runs~\cite{darnton2007,son2013}. In addition, hydrodynamic interactions with solid boundaries allow them to perform complex trajectories, such as circles~\cite{lauga2006,shenoy2007}. This is not the case for most artificial microswimmers, e.g. self-propelled colloids, whose orientational motion is fully determined by rotational diffusion under homogeneous conditions, thus performing a persistent random walk~\cite{howse2007,sano2010,palacci2010}. In order for an active colloid to exhibit persistent rotations, external torques can be applied by e.g. magnetic fields~\cite{baraban2013}. Moreover, 
chirality can induce a coupling between the rotational and translational motion of a microswimmer, which enables self-rotation~\cite{loewen2016}. This can be realized by using molecularly-chiral materials, such as liquid crystal droplets, which exhibit helical auto-propulsion in aqueous surfactants~\cite{krueger2016,yamamoto2017,yamamoto2018}.
Chirality is also possible for an asymmetric active colloid, in such a way that it can undergo a velocity-dependent viscous torque. This results in circular orbits whose direction and radius are simply determined by the specific particle geometry~\cite{nakata1997,takagi2013, kuemmel2013,wykes2016}.

In this Letter, we experimentally demonstrate that  even under uniform conditions, rigid spherical active colloids can experience a transition from diffusive rotational motion to persistent circular orbits above a critical propulsion speed when moving in homogeneous viscoelastic fluids. Unlike chiral microswimmers, here the circular trajectories  can spontaneously reverse their direction, with angular speed and radius of curvature that non-linearly depend on the velocity.  Such an intricate behavior can be captured by a minimal 2D Langevin model with memory, where a non-linear coupling between orientational and translational motion supplemented by thermal noise reproduces the aforementioned transition as well as the spontaneous changes in sign of the orbits.

\begin{figure*}
	\includegraphics[width=1.7\columnwidth]{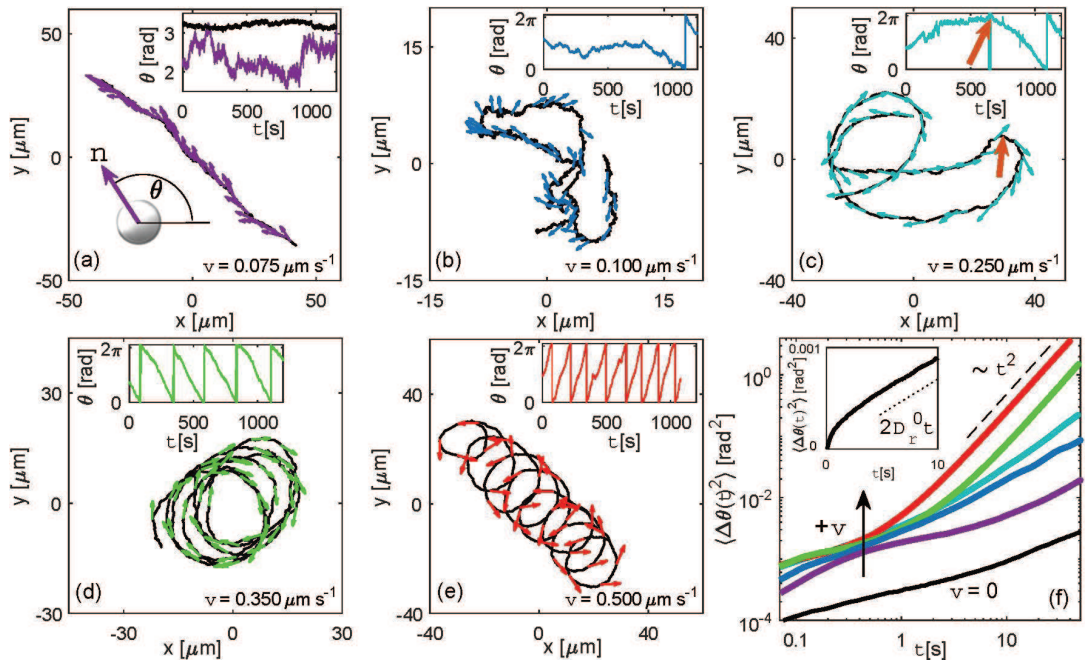}
\caption{(Color online) (a) Image of a spherical colloid, which is half-coated with carbon (dark area). The arrow pointing from the capped to the uncapped hemisphere represents its orientation $\mathrm{\textbf{n}}$, which defines the angle $\theta$ with respect to the x-axis. Some exemplary trajectories $\mathrm{\textbf{r}} = (x,y)$ (solid lines) and orientations $\mathrm{\textbf{n}} = (\cos \theta, \sin \theta)$ (arrows) at different propulsion speeds are plotted in: (a) $v=0.075\,\mu\mathrm{m}\,\mathrm{s}^{-1}$, (b) $v=0.100\,\mu\mathrm{m}\,\mathrm{s}^{-1}$, (c) $v=0.250\,\mu\mathrm{m}\,\mathrm{s}^{-1}$, (d) $v=0.350\,\mu\mathrm{m}\,\mathrm{s}^{-1}$, and (e) $v=0.500\,\mu\mathrm{m}\,\mathrm{s}^{-1}$. Insets: time evolution of the corresponding angle $\theta$. The inset of Fig.~\ref{fig:fig1}(a) also shows the time evolution of $\theta$ for a passive colloid ($v=0$, upper curve). (f) Log-log representation of the mean-squared angular displacements for different propulsion speeds, increasing from bottom to top. Inset: expanded view of the mean-squared angular displacement for a passive colloid, showing the change from subdiffusive to diffusive behavior with increasing $t$. The dotted and dashed lines are guides to the eye to indicate diffusive and ballistic behavior, respectively. }
\label{fig:fig1}
\end{figure*}

In our experiments, we use as synthetic microswimmers spherical silica particles (radius $a = 3.88\,\mu\mathrm{m}$) half-coated with carbon and suspended in a viscoelastic binary fluid kept at $T = 296$~K. Such a  fluid is composed of water and propylene glycol n-propyl ether (PnP), with added polyacrylamide (PAAm) polymers in the semidilute regime, the concentration of which determines its viscoelasticity. The resulting viscosities are 2 orders of magnitude higher than that of water while the relaxation times are several seconds, thereby leading to strong memory effects in the motion of the microswimmers. Their self-propulsion is achieved within a quasi-2D cell by local demixing of the binary fluid, which is induced by laser-heating of their asymmetrically light-absorbing surface \cite{samin2015,gomezsolano2017}. The propulsion speed, $v$, is controlled by the laser intensity $I$. For the intensities considered here, $v \propto I$, resulting in an active motion directed away from the carbon cap. A detailed description of our experimental setup is provided in~\cite{gomezsolano2017,gomezsolano2016}.

In Figs.~\ref{fig:fig1}(a)-(e) we plot some representative 2D trajectories $\mathrm{\textbf{r}} = (x,y)$ of a self-propelled colloid moving in the PAAm solution at $0.05$~wt~\%, at different  $v$, as well as the instantaneous orientation $\mathrm{\textbf{n}} = (\cos \theta, \sin \theta)$. The latter is defined as the unit vector pointing from the capped to the uncapped hemisphere, as depicted in Fig.~\ref{fig:fig1}(a). In the insets of Figs.~\ref{fig:fig1}(a)-(e) we also show the time evolution of the angular coordinate $\theta$, from which we compute the corresponding mean-squared angular displacements $\langle \Delta \theta(t)^2 \rangle$, see Fig.~\ref{fig:fig1}(f). For a passive colloid ($v=0$), the angular motion is subdiffusive at short time-scales due to the viscoelasticity of the surrounding fluid~\cite{andabloreyes2005}, as shown in the inset of Fig.~\ref{fig:fig1}(f). Angular diffusion is observed at long-time scales, i.e.  $\langle \Delta \theta(t)^2 \rangle = 2D_r^0 t$, see inset of Fig.~\ref{fig:fig1}(f), with a diffusion coefficient $D_r^0 = k_B T/(8\pi a^3\eta_0)$ which satisfies the Stokes-Einstein relation, where $\eta_0$ is the zero-shear viscosity of the fluid. 
Interestingly, we find that at small $v$, a similar angular diffusive behavior persists, i.e. $\langle \Delta \theta(t)^2 \rangle = 2D_r t$ at sufficiently long times, as shown in Fig.~\ref{fig:fig1}(f). Nevertheless, under such conditions the effective rotational diffusion coefficient $D_r$ is higher than $D_r^0$, in agreement with previous observations~\cite{gomezsolano2016}. For instance, at $v = 0.100\,\mu\mathrm{m}\,\mathrm{s}^{-1}$ the trajectories are rather wavy, see  Fig.~\ref{fig:fig1}(b), where $D_r$ is 2 orders of magnitude higher than $D_r^0$.
Remarkably, a further increase of $v$ leads to an unexpected transition to a different dynamical regime characterized by a persistent rotation. In such a case, the time evolution of $\theta$ is on average linear over intervals of several minutes, as shown in  the insets of Figs.~\ref{fig:fig1}(c)-(e). Such orbits can spontaneously reverse their direction, as observed in  Fig.~\ref{fig:fig1}(c) at $v = 0.250\,\mu\mathrm{m} \, \mathrm{s}^{-1}$, where an intially counter-clockwise rotation reverses after 700 seconds (thick arrows). Furthermore, the circular trajectories become better defined with increasing $v$, as shown in  Figs.~\ref{fig:fig1}(d)-(e), where the rotational reversal becomes very uncommon over the measurement time.
In this regime, at sufficiently long time-scales the mean-squared angular displacement behaves as  $\langle \Delta \theta(t)^2 \rangle =2D_r t + \omega^2 t^2$, see  Fig.~\ref{fig:fig1}(f). 
This allows to characterize the motion for cycles with the same sense of rotation by the angular velocity $\omega$.

\begin{figure}
\includegraphics[width=0.5\textwidth]{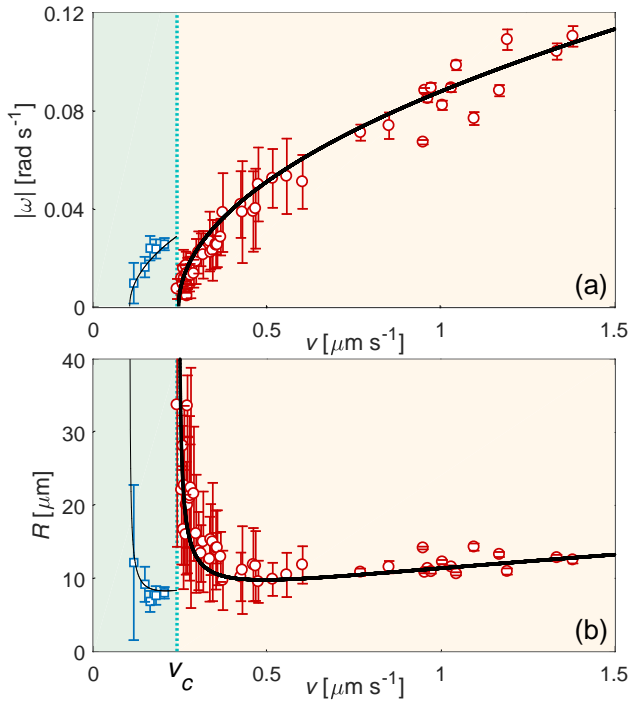}
\caption{(Color online) (a) Dependence of the angular speed $|\omega|$ of the circular orbits on the propulsion speed $v$ for an active colloid moving in a PAAm solution at 0.05~wt~\% ($\circ$) and 0.06~wt~\% ($\square$). The vertical dotted line depicts the critical velocity $v_c$ in the former case, below which only enhanced rotational diffusion occurs. (b) Dependence of the corresponding radius of curvature $R$ on $v$, same symbols as in \ref{fig:fig2}(a). The solid lines in \ref{fig:fig2}(a) and \ref{fig:fig2}(b) represent the dependence on $v$ given by Eq.~(\ref{eq:omega}), while the error bars correspond to the standard deviation over several cycles with the same sense of rotation.}
\label{fig:fig2}
\end{figure}

In Fig.~\ref{fig:fig2}(a) we plot the angular velocity $\omega$ as a function of $v$. We point out that both directions, i.e. clockwise ($+$) and counter-clockwise ($-$), are possible for the same particle, as illustrated in Figs.~\ref{fig:fig1}(d) and~\ref{fig:fig1}(e), respectively. Furthermore, we find that $|\omega|$ depends nonlinearly on $v$. While below a threshold ($v_c = 0.240 \, \mu\mathrm{m}\,\mathrm{s}^{-1}$) only enhanced angular diffusion occurs, $|\omega|$ exhibits an approximately square-root growth with $v$ above $v_c$. The propulsion-speed threshold and the angular velocity of such orbits strongly depend on the polymer concentration. For instance, at higher PAAm concentration (0.06~wt~\%), the transition occurs at smaller propulsion speed ($v_c = 0.105 \, \mu\mathrm{m}\,\mathrm{s}^{-1}$), as demonstrated in Fig~\ref{fig:fig2} (a). In addition, the radius of curvature of the orbits, defined as $R = v/|\omega|$, displays a non-monotonic dependence on $v$, as shown in Fig.~\ref{fig:fig2}(b), where a very sharp decrease of $R$ just above $v_c$ is followed by a weak variation at higher propulsion speeds.  
We point out that such a behavior significantly differs from that of chiral active particles, e.g. L-shaped colloids~\cite{kuemmel2013}, for which orbiting trajectories result from a velocity-dependent torque due to viscous forces exerted at a non-zero lever arm \cite{tenhagen2015}. In such a case, a persistent rotation occurs even at vanishingly small $v$, where $\omega$ increases linearly with $v$, while $R$ remains constant as it is only determined by the particle geometry. Besides, the sign of $\omega$ is fixed by the chirality of the particle, unlike the spontaneous rotational reversal observed in the viscoelastic fluid. Thus, our findings suggest that the emergence of circular motion for the spherical active colloids is due to the response of the surrounding fluid rather than the particle chiral asymmetry~\cite{footnote1}.

As a matter of fact, our results can be rationalized by a phenomenological model which takes into account the memory of the system due to the viscoelasticity of the fluid environment. A stress-relaxation modulus which mimics the mechanical response of the fluid is $G(t) =  2\eta_{\infty} \delta(t) + \frac{\eta_0 - \eta_{\infty}}{\tau}e^{-\frac{t}{\tau }}$~\cite{bird1987,paul2018}.
Here, the first term accounts for the instantaneous relaxation of the solvent with viscosity $\eta_{\infty}$, whereas the second term captures the time-delayed response of the polymer solution with relaxation time $\tau$ and zero-shear viscosity $\eta_0$. The coupling with the translation of the colloidal particle is included by the memory friction kernel $\Gamma_T(t) =6\pi a G(t) $. 
Since the system is overdamped and without external forces, we describe the 2D motion of the particle position $\mathrm{\textbf{r}} = (x,y)$ by the non-Markovian Langevin equation~\cite{mori1965}
\begin{equation}\label{eq:Langevinposition}
-\int_{-\infty}^t \Gamma_T(t - t')\left[ \dot{\mathrm{{\bf {r}}}}(t') - \mathrm{{\bf {v}}}(t')\right]\mathrm{d}t'  + {\bm{\zeta}}_{T}(t) = \bm{0}.
\end{equation}

The term $\mathrm{\textbf{F}}_{H}(t)=-\int_{-\infty}^t \Gamma_T(t - t')\left[ \dot{\mathrm{{\bf {r}}}}(t') - \mathrm{{\bf {v}}}(t')\right]\mathrm{d}t'$ represents the total hydrodynamic force at time $t$ exerted by the fluid on the particle, moving at propulsion velocity $\mathrm{{\bf {v}}}(t')= v\mathrm{\textbf{n}}(t')$, i.e.  parallel to the particle orientation, as experimentally observed. The term $\bm{\zeta}_T(t)$ is a zero-mean Gaussian noise, which mimics thermal fluctuations. For the sake of simplicity, we assume that it satisfies the second fluctuation-dissipation theorem~\cite{medinanoyola1987}, i.e. $\langle  \zeta_T^i(t) \zeta_T^j(s) \rangle= k_B T \delta_{ij}\Gamma_T(|t - s|)$. Note that the total hydrodynamic force vanishes on average, $\langle \mathrm{\textbf{F}}_{H}(t) \rangle = - \langle  {\bm{\zeta}}_{T}(t) \rangle = \bm{0}$, thus fulfilling the force-free condition. 
By analogy with active Brownian motion in Newtonian liquids~\cite{tenhagen2015,tenhagen2011}, in Eq.~(\ref{eq:Langevinposition}) we can interpret the term $\mathrm{\textbf{F}}_{v}(t)=v\int_{-\infty}^t \Gamma_T(t - t')\mathrm{{\bf {n}}}(t')\mathrm{d}t'$ as an internal force related to the active process of self-propulsion~\cite{takatori,yan}. We point out that, due to the memory of the system, $\mathrm{\textbf{F}}_{v}(t)$ is not parallel to the instantaneous particle orientation $\mathrm{{\bf {n}}}(t)$, as it depends on all previous times $t' \le t$. Therefore, in general $\mathrm{\textbf{F}}_{v}(t)$ lags behind $\mathrm{\textbf{n}}(t)$. Such a time-delayed response suggests a coupling between $\mathrm{\textbf{F}}_{v}(t)$ and  $\mathrm{{\bf {n}}}(t)$ by means of an internal torque $\mathbf{T}_v(t) = {T}_v(t)\hat{\mathbf{z}} = \mathrm{\textbf{L}}(t) \times \mathrm{\textbf{F}}_{v}(t)$, where $\hat{\mathbf{z}} = \hat{\mathbf{x}} \times \hat{\mathbf{y}}$ is the unit vector orthogonal to the $\hat{\mathbf{x}}$ and $\hat{\mathbf{y}}$ directions. Here, the effective lever arm $\mathrm{\textbf{L}}(t) = -\mu a \mathrm{\textbf{n}}(t)$ mimics the spatial delay of $\mathrm{\textbf{F}}_{v}(t)$ with respect to the particle center, where we expect $\mu \sim O(1)$. $T_v$ consistently vanishes in the case of passive motion ($v=0$) and for active motion in Newtonian fluids ($\eta_0 = \eta_{\infty}$).

On the other hand, $\mathrm{{\bf {n}}}(t)$  is affected by the viscoelasticity of the surrounding fluid, as demonstrated by the subdiffusive regime at short times in Fig.~\ref{fig:fig1}(f). 
Therefore, apart from the effect of $\mathbf{T}_v(t)$, we consider that $\mathrm{\textbf{n}}$ is also subjected both to an angular drag with memory kernel $\Gamma_R(t) ={ 8\pi a^3} G(t)$ and to zero-mean thermal noise $\zeta_R$, which satisfies~\cite{andabloreyes2005}: $\langle  \zeta_R(t) \zeta_R(s) \rangle= k_B T \Gamma_R (|t - s|)$.
Consequently, we model its 2D orientation by the non-Markovian equation
\begin{equation}\label{eq:Langevinangle}
	-\int_{-\infty}^t \Gamma_R(t - t')\dot{\theta}(t')\mathrm{d}t' +  T_v(t) + \zeta_R(t) = 0.
\end{equation}

\begin{figure}
\includegraphics[width=0.475\textwidth]{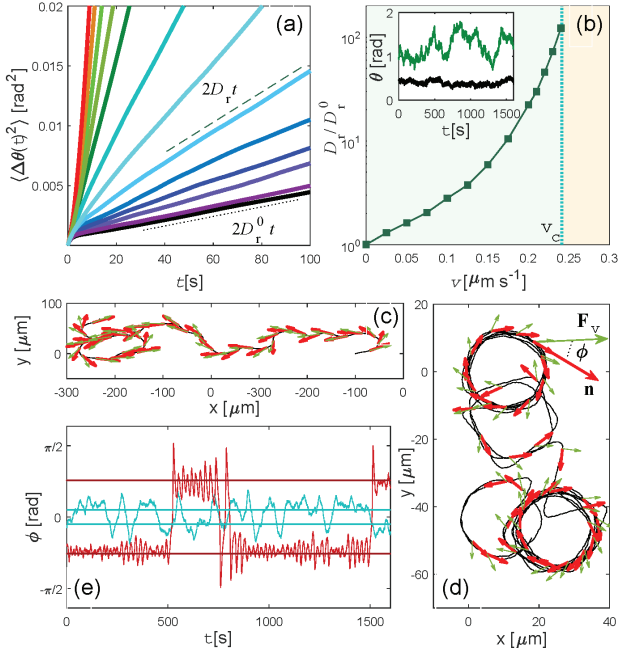}
\caption{(Color online) (a) Mean-squared angular displacements computed from numerical solutions of Eqs.~(\ref{eq:Langevinposition})-(\ref{eq:Langevinangle}) at different propulsion speeds $v < v_c = 0.240\,\mu\mathrm{m}\,\mathrm{s}^{-1}$, increasing from bottom to top. (b) Corresponding effective rotational diffusion coefficient as a function $v$. The vertical dashed line represents $v_c$. Inset: Time evolution of $\theta$ for a numerical passive trajectory ($v=0$, bottom) and an active one ($v= 0.200 \,\mu\mathrm{m}\,\mathrm{s}^{-1}$, top). Examples of numerical trajectories at different propulsion speeds above $v_c$: (c) $v = 0.250 \,\mu\mathrm{m}\,\mathrm{s}^{-1}$, and (d) $v =  0.600 \,\mu\mathrm{m}\,\mathrm{s}^{-1}$. The thick arrows represent the orientation $\mathrm{\textbf{n}}$, whereas the thin arrows show the corresponding propulsion force $\mathrm{\bf{F}}_v$. (e) Time evolution of the phase difference between $\mathrm{\textbf{n}}$ and $\mathrm{\bf{F}}_v$ for a particle moving at $v = 0.250 \,\mu\mathrm{m}\,\mathrm{s}^{-1}$ (inner curve) and at $v = 0.600 \,\mu\mathrm{m}\,\mathrm{s}^{-1}$ (outer curve). The solid lines represent the steady-state values $\phi_{\pm}$.}
\label{fig:fig3}
\end{figure}

The minimal model (\ref{eq:Langevinposition})-(\ref{eq:Langevinangle}) reproduces the main non-trivial features of the active motion in the viscoelastic fluid~\cite{footnote2}. First, at small $v$, numerical solutions of Eqs.~(\ref{eq:Langevinposition}) and (\ref{eq:Langevinangle}) evidence diffusive orientational dynamics at sufficiently large time-scales, i.e. $t \gg \frac{\eta_{\infty}}{\eta_0}\tau$, as shown in~Fig.~\ref{fig:fig3}(a), with an effective rotational diffusion coefficient $D_r$ that significantly grows with increasing $v$, see Fig.~\ref{fig:fig3}(b). At small $v$, the nonlinear term  $T_v $ in Eq.~(\ref{eq:Langevinangle}) leads to short-lived rotations, which are  randomized by thermal fluctuations, thereby resulting in an enhanced angular motion, as observed both in experiments and simulations, see insets of Figs.~\ref{fig:fig1}(a) and~\ref{fig:fig3}(b), respectively. On the other hand, around the critical propulsion speed $v_c = \frac{4a}{3\mu\tau\left( 1 - \frac{\eta_{\infty}}{\eta_0} \right)}$,
a maximum in $D_r$ is achieved, while at $v > v_c$, circular orbits become stable solutions of the deterministic parts of Eqs~(\ref{eq:Langevinposition}) and (\ref{eq:Langevinangle}), with a persistent angular velocity
\begin{equation}\label{eq:omega} 
	\omega = \pm \frac{1}{\tau} \sqrt{\frac{v}{v_c} - 1}.
\end{equation}

The transition from active-Brownian to circular motion when increasing $v$ from below to above $v_c$ is illustrated in Figs~\ref{fig:fig3}(c)-(d), where we plot typical trajectories obtained by numerically solving Eqs~(\ref{eq:Langevinposition})-(\ref{eq:Langevinangle}). Note that both $v_c$ and $|\omega|$ strongly  depend on the viscoelasticity of the fluid, simplified by the parameters $\tau$, $\eta_0$ and $\eta_{\infty}$. Eq.~(\ref{eq:omega}) displays both rotational directions, with a square-root dependence on $v$, in excellent agreement with our experimental findings, as shown as solid lines in Fig~\ref{fig:fig2}(a). Here, the only fitting parameter is the dimensionless lever-arm $\mu$, which is set to $\mu =1.3$ for both PAAm concentrations in order to get self-consistent agreement between the experimental data, the predicted values of $v_c$, $\omega$, $R$ and the numerical results. 
In Fig.~\ref{fig:fig2}(b) we demonstrate that the radius of curvature can also be well described by $R=v\tau \left(\frac{v}{v_c}-1\right)^{-\frac{1}{2}}$ (solid lines). While $R$ abruptly decreases just above $v_c$, after a minimum value $R=2v_c\tau$ at $v = 2v_c$ it exhibits a subsequent slow increase at larger $v$. In addition, numerical solutions at $v > v_c$, as those depicted in Figs.~\ref{fig:fig3}(c) and (d), show that the circular trajectories can switch their angular velocity between $+|\omega|$ and $-|\omega|$ for sufficiently long times. Such orientational flips occur when a thermal fluctuation is comparable to the mean torque $8\pi \eta_0 a^3 |\omega|$ in order to destabilize the current orbit. We also check that, unlike the quasi-instantaneous response of Newtonian fluids to self-propulsion~\cite{gomezsolano2017}, here $\mathrm{\bf{F}}_v(t)$ significantly lags behind $\mathrm{\textbf{n}}(t)$. This is highlighted in the trajectories of Fig~\ref{fig:fig3}(c)-(d), where the thin and thick arrows represent instantaneous values of $\mathrm{\bf{F}}_v(t)$ and  $\mathrm{\textbf{n}}(t)$, respectively. Their phase difference $\phi$, defined by the large arrows in Fig~\ref{fig:fig3}(d), exhibits less and less stochastic jumps between the two steady-state values $ \phi_{\pm} = \pm \arctan\left( \frac{\eta_0 - \eta_{\infty}}{\eta_0 +\eta_{\infty}\omega^2\tau^2} |\omega | \tau \right)$  with increasing $v$, see Fig~\ref{fig:fig3}(e). This translates into stable circular orbits at sufficiently high propulsion speed,  where the spontaneous reversals of the sense of rotation become very infrequent, as experimentally observed.

In conclusion, we have investigated the active motion of spherical colloids in a viscoelastic fluid over a broad velocity range. Our findings uncover a transition from a persistent random walk with enhanced angular diffusion to circular motion with stochastic orientational flips. Unlike synthetic microswimmers with molecular or shape chirality, here the emergence of such circular orbits is due to the delayed response of the surrounding fluid, as explained by a minimal non-Markovian Langevin model.
Although this phenomenological description includes neither the detailed flow field around the microswimmer~\cite{natale2017,datt2017}, nor a possible modification of the noise and friction due to the non-equilibrium particle motion~\cite{maes2014,krueger2017}, it reproduces the main non-trivial features observed in the experiments. Therefore, our findings are expected to be robust to such details and to happen for active colloids in other kinds of viscoelastic media~\cite{gomezsolano2013,gomezsolano2015}, as well as for deformable  microswimmers~\cite{ohta2009,moerman2017,vutukuri2017}. Moreover, it will be important to explore memory-induced effects for microswimmers  in viscoelastic media under other experimental conditions, e.g. gradients~\cite{gomezsolano2017}, corrugated substrates~\cite{gomez2015,choudhury2017}, time-dependent motility~\cite{lozano2018} and moving  potentials~\cite{berner2018}, as they are commonly encountered in nature and technological applications.

This work was supported by the German Research Foundation (DFG) through grant No. GO 2797/1–1 (J.R.G.S.), by the DFG through the priority programme SPP 1726 on microswimmers (C.B.), by the ERC Advanced Grant ASCIR grant No. 693683 (C.B.).

\end{document}